# Highly wear-resistant and low-friction $Si_3N_4$ composites by addition of graphene nanoplatelets approaching the 2D limit

*Orsolya Tapasztó[1], Jan Balko[2], Viktor Puchy[2], Péter Kun[1], Gergely Dobrik[1], Zsolt Fogarassy[1], Zsolt Endre Horváth[1], Jan Dusza[2], Katalin Balázsi[1], Csaba Balázsi[1], Levente Tapasztó[1*]*

1. Institute of Technical Physics and Materials Science, Centre for Energy Research, Hungarian Academy of Sciences, Konkoly Thege str. 29-33, 1121 Budapest, Hungary
2. Institute of Materials Research, Slovak Academy of Sciences, Watsonova 47, 040 01 Košice, Slovak Republic

## Abstract

Graphene nanoplatelets (GNPs) have emerged as one of the most promising filler materials for improving the tribological performance of ceramic composites due to their outstanding solid lubricant properties as well as mechanical and thermal stability. Yet, the addition of GNPs has so far provided only a very limited improvement in the tribological properties of ceramics, particularly concerning the reduction of their friction coefficient. This is most likely due to the challenges of achieving a lubricating and protecting tribo-film through a high GNP coverage of the exposed surfaces. Here we show that this can be achieved by efficiently increasing the exfoliation degree of GNPs down to the few-layer (FL) range. By employing FL-GNPs as filler material, the wear resistance of $Si_3N_4$ composites can be increased by about twenty times, the friction coefficient reduced to nearly its half, while the other mechanical properties are also preserved or improved. Using confocal Raman microscopy, we were able to demonstrate the formation of a continuous FL- GNP tribo-film, already at 5wt% FL-GNP content.

*Corresponding author. Tel: +36-1-3922680. E-mail: tapaszto@mfa.kfki.hu (Levente Tapaszto)

# 1. Introduction

The outstanding wear resistance of ceramic materials is exploited in a wide range of technological applications. For contact mechanical applications (e.g. bearings, valves) the high wear resistance is desirable to be also associated with a low friction coefficient. Significantly decreasing the friction coefficient is essential for reducing the losses in moving and rotating parts, while increasing the wear resistance improves the durability and lifetime of the components. It is also particularly important, that these benefits come without compromising the other mechanical properties such as toughness, hardness or flexural strength of the composite. The addition of layered materials, known as excellent solid lubricants, is a promising strategy to improve both the mechanical and tribological performance of ceramics. Graphene nanoplatelets have emerged as a particularly promising nanoscale filler phase for various composite materials due to their exceptional mechanical properties and thermal stability[1]. They also hold the potential for improving the tribological properties of the composites as graphite is known to be an excellent solid lubricant[2].

The research conducted so far on the effect of GNP addition on the tribological properties of the ceramics indicates that the results are strongly dependent on the type of the ceramic matrix. A significant increase (about an order of magnitude) in the wear resistance has been reported for $Al_2O_3$[3] and $SiO_2$ matrix composites[4]. By contrast, for other technologically relevant ceramic materials, such as SiC[5], $Si_3N_4$[6] and $ZrO_2$ composites[7] only a moderate 50-70% improvement of the wear rate could be achieved. Concerning the reduction of the friction coefficient, the GNP addition so far has proven even less effective, in several cases the stationary friction coefficient of the ceramic/GNP composites increased or the reduction was only marginal. The most promising results again have been reported for $Al_2O_3$ where a friction coefficient reduction of up to 20-30% could be achieved[3].

In the present work we focus on the tribological properties of $Si_3N_4$/GNP composites. Although the addition of GNPs has brought spectacular results in improving the electrical conductivity[8] and enhancing the fracture toughness[9,10] of $Si_3N_4$ composites, improving the tribological properties has proven particularly challenging. Belmonte *et al.* reported the improvement of the wear resistance up to 56% for $Si_3N_4$ ceramics prepared by spark plasma sintering (SPS) with 3 wt% GNP addition[11]. Due to the GNP addition, the friction coefficient increased for loads below 100N, while a slight reduction of friction could be observed at 200 N, under isooctane lubricated conditions. The elastic modulus, hardness, and fracture toughness of the 3wt% GNPs composites has slightly decreased as compared to the monolithic material. For $Si_3N_4$ samples sintered by hot isostatic pressing (HIP) with 3wt% of GNP addition Hvizdos et al. reported a 60% increase of the wear resistance under dry sliding conditions; however, the friction coefficient remained practically unchanged[6]. Hot pressed $Si_3N_4$/GNP composites with up to 10wt% GNP content investigated by Rutkowski et al. displayed an increasing wear rate and friction coefficient with GNP content above 2wt%, while the other mechanical properties of the composites have also declined[12]. Recently, a detailed study has been conducted with our participation on the tribological properties of the $Si_3N_4$/GNP composites, prepared by both SPS and HIP methods[13]. Only a limited improvement of the tribological properties could be achieved. Our conclusion was that this is most likely due to lack of formation of a GNP-based lubricant film on the composite surfaces. Recently $Si_3N_4$ composites with 2wt% of boron nitride nanoplatelets addition have been prepared by hot pressing[14]. Such composites displayed about 20% of reduction in the friction coefficient as compared to monolithic $Si_3N_4$.

The picture unfolding from the results concerning the tribological performance of the $Si_3N_4$/GNP composites is a moderate improvement of the wear resistance while the friction coefficient is often increased or only marginally reduced. This is in striking contrast with the excellent solid lubricant properties of graphite, which are expected to be inherited also by GNPs. It has been shown that the lubricating properties of graphite are preserved down to four layers, after which the friction gradually increases as the number of layers is further reduced[3]. As the usually applied GNPs have a thickness of orders of tens of graphene layers, this is clearly not a limiting factor. However, Scanning Electron Microscopy (SEM) and Raman analysis of the wear tracks revealed the lack of a continuous GNP tribo-film formation[6,11,13] that could be at the origin of the limited improvement in the tribological performance. The reason for this is the relatively low GNP coverage of the contact surfaces. This is surprising, as graphene is characterized by an enormous surface area up to 2630 $m^2$/g in the single layer limit. This means that already a few percent of graphene content, if homogeneously distributed within the volume of the composite matrix, easily enables a quasi-continuous coverage of practically any composite surface. The fact that the continuous GNP coverage is still difficult to achieve is most likely due to the incomplete nature of exfoliation yielding a thickness distribution far from the two dimensional limit, and a much lower surface to volume ratio of the GNPs. A straightforward solution to this problem is increasing the exfoliation degree of the GNPs.

Various exfoliation methods have been employed aiming to produce large quantities of single and few layer graphene flakes with their structure kept as intact as possible. The chemical route is highly efficient in improving the exfoliation degree of graphite[15]. However, it provides graphene oxide layers, which have been found to be far less effective in improving the mechanical and tribological properties of ceramic composites, even in their reduced form, as compared to chemically unaltered graphene[3,5]. The liquid phase exfoliation of graphite[16] is a promising route towards obtaining few layer graphene sheets with noncovalent functionalization; however the yield of thick multilayer flakes is relatively high. Here we employ an even simpler mechanochemical exfoliation method based on the ball milling of graphite with melamine addition to increase the exfoliation efficiency. The efficiency of the melamine has been attributed to the fact that it can easily intercalate graphite and form large 2D networks in-between the graphite layers, weakening the van der Waals forces between them that in turn allows an easy exfoliation upon mild shear stress during ball milling[17,18]. Although, the method has been originally developed to provide stable water dispersions of few-layer graphene sheets, we show that FL-GNPs produced this way, can be highly efficient filler materials for ceramic composites, especially regarding the improvement of the tribological properties.

## 2. Experimental

The graphene nanoplatelets have been exfoliated from commercial graphite powder (Aldrich) by attritor milling (Union Process, type 01-HD/HDDM) for 10h at 3000 rpm in ethanol[19]. After milling, 30 mg of the as-prepared GNPs were introduced in a planetary mill ($ZrO_2$ balls), with 90 mg of commercially available melamine (Aldrich) addition, and ball-milled at 225 rpm for 30 min. The resulting solid mixtures were dispersed in 20 ml water and sonicated for 30 min. The obtained FL-GNPs have been investigated by a Philips CM-20 Transmission electron microscope (TEM) at 200 keV and Raman spectroscopy (Witec Alpha 300 RSA).

For silicon nitride preparation, 90 wt% $Si_3N_4$ (Ube, SN-ESP) has been used as starting powder, while as sintering aids we used 4 wt% $Al_2O_3$ (Alcoa, A16) and 6 wt% $Y_2O_3$ (H.C. Starck, grade C) powders. The powder mixtures were milled in ethanol using an attritor mill (Union Process, type 01-HD/HDDM) at 4000 rpm for 5h, in a silicon nitride tank, using $ZrO_2$ balls. 3 and 5 wt% of exfoliated FL-GNPs (preparation process described above) have been added to the batches and homogenized by mechanical milling at 600 rpm for 30 minutes. The as-prepared powder mixtures have been subjected to spark plasma sintering (SPS, HD P5, FCT - SYSTEME GMBH, GERMANY). The mixtures were sintered into discs of 3 cm in diameter and 5 mm thickness, at 1700°C in vacuum, with 10min dwell time. The powders were heated with 100°C min$^{-1}$ heating rate for all the experiments using on/off current pulses of 12/2, 5500 A and 7,2 V. An applied uniaxial compression of 50 MPa and chamber pressure of 1 mbar has been maintained during sintering. The microstructural characterization of the sintered samples has been carried out by a LEO (ZEISS) 1540 XB field emission scanning electron microscope and XRD measurements (Bruker AXS D8). The wear behaviour of the materials was studied by an unlubricated ball-on-disc setup (ASTM G99-03, Standard test method for wear testing with a pin-on-disk apparatus, 2003) using a tribometer (DTHT 70010, CSM Instrument, Switzerland), against silicon nitride balls (6 mm diameter) with a track radius of 2.5 mm. The tested sample surfaces were ground and polished with a final diamond suspension of 3 μm (roughness Ra<0.25 μm). The specimens were tested normal to the major surface of the sample with an applied load of 5 N. The tests were performed in air, at room temperature with a relative humidity of 40±5%. The total sliding distance was 300 m and the sliding velocity was 10cm/s. The coefficient of friction was continually recorded during the tests. The wear volume of each specimen was calculated from the surface profile traces (at least 6) across the wear track and perpendicular to the sliding direction using a profilometer (Mitutoyo SJ-201, USA), and by using a high precision confocal microscope (PLuneox 3D Optical Profilometer, Sensofar, Spain). The wear tracks have also been investigated by Raman spectroscopy, using a confocal Raman microscope system (Witec Alpha 300 RSA) with Nd:YAG laser (532 nm, 20 mW). For Raman mapping 2500 spectra have been acquired with 600 lines/mm grating from a 50 μm-wide rectangular area with 1 μm spacing. The bending strength of the composites has been determined by three-point bending tests on a tensile/loading machine (INSTRON-1112). The hardness and fracture toughness values have been obtained from micro-indentation experiments on a hardness tester (KS Prüftechnik) applying a load of 10Kp for 10 seconds. The fracture toughness was calculated according to the Anstis-formula[20], using the Young's modulus value of $E$ = 250 GPa. For the three-point bending strength measurements the samples have been cut into rectangular specimens of 3.5 mm x 5 mm x 28 mm.

## 3. Results and Discussions

To investigate the efficiency of the applied exfoliation method and the structure of the resulting GNPs we have performed TEM measurements on GNP samples prepared by the very same exfoliation process (see experimental section) both with and without melamine addition. Two characteristic TEM images are shown in Figure 1. From the TEM investigations it becomes evident that the GNPs prepared by melamine addition are significantly thinner than the conventionally exfoliated GNPs. Based on TEM investigations, we estimate that the presence of melamine during milling reduces the average platelet thickness below 10 layers. This is in full accordance with the literature for melamine enhanced exfoliation of graphite[17,18]. By contrast, conventionally prepared GNPs typically consist of 40-60 layers. Nevertheless, even in the case of melamine addition we only approach but typically do not reach the 2D limit, namely single graphene layers. However, this might be an advantage for tribological applications, as platelets below 4 layers are characterized by diminished lubricant properties[15]. The advantage of using melamine is that it is noninvasive, it does not form strong chemical bonds to the graphene scaffold, that would modify its electrical and mechanical properties[17]. We confirm this by Raman spectroscopy investigations of the resulting FL-GNPs that display a relatively small D peak (mainly attributed to flake edges), indicative of high structural quality of the FL-GNPs. After the exfoliation by milling, the melamine can be removed using hot water, leaving behind the pure FL-GNP powder. Potential melamine residues are fully removed during the sintering process at 1600°C, as melamine decomposes over 350°C [21].

The microstructure of the sintered $Si_3N_4$/FL-GNP composites has been investigated by Scanning Electron Microscopy and XRD. The structural data obtained from XRD measurements (not shown) are summarized in Table 1. Lower magnification SEM images (Fig.2a) display a relatively high density of darker areas with micron scale characteristic lateral size. Higher magnifications (Fig. 2b) reveal that these areas correspond to FL-GNPs laying on the surface of the $Si_3N_4$ composite. Form higher resolution SEM images, FL-GNPs both parallel and perpendicular to the surface can be identified, which is in contrast to the SEM images most often reported in the literature, where usually only a few FL-GNPs perpendicular to the surface can be identified. We attribute this to the much higher surface area covered by FL-GNPs as compared to their thicker GNP counterparts. It is also known that sintering processes employing uniaxial pressure yield GNPs with a preferential orientation perpendicular to the pressure axis[22]. Nevertheless, we observed a high concentration of GNPs in SEM images acquired on various fracture surfaces. It is also worth noting that in SEM images it is much easier to identify the thicker FL-GNPs that provide a stronger contrast, while atomically thin layers are hard to discern. The picture unfolding from the TEM and SEM analysis is a large density of thin FL-GNP covering all sample surfaces, which is quite promising for the tribological performance of such composites.

Figure 3 shows the friction coefficient of various composite samples under the same experimental conditions, acquired using a $Si_3N_4$ ball (D = 6 mm) under dry sliding conditions, with 5N loading and 10 cm/s sliding speed. The addition of 3wt% of conventional GNP exfoliated without melamine addition results in a friction coefficient of 0.79 +/− 0.02, practically identical to that of the monolithic $Si_3N_4$. This is consistent with our previous results, as well as the reports from the literature [6,11]. The addition of 3 wt % of FL-GNPs exfoliated with melamine addition results in a roughly 15% reduction of the friction coefficient. However, the most exciting results have been obtained for 5 wt % of FL-GNP addition, where the steady state friction coefficient was found 0.43 +/− 0.04, almost half of the value for monolithic $Si_3N_4$. To our knowledge such a substantial reduction of the friction coefficient upon GNP addition has not been reported before in $Si_3N_4$ or any other ceramic material. It is also worth noting that in contrast to monolithic and 3 wt% GNP samples, for samples with FL-GNP addition the friction coefficient slightly but steadily decreases as a function of sliding distance. This might indicate the self-lubricating nature of the composite leading to the exposure (pulling out) of more and more FL-GNP flakes, which, together with the high initial FL-GNP surface coverage, further improve the lubricating properties. The optical images of the wear track after 300m sliding have been investigated by optical microscopy. Fig. 4 displays two representative optical images for the monolithic $Si_3N_4$ and the 5 wt %FL-GNP addition. For the monolithic $Si_3N_4$ sample a high density of abrasion groves can be identified, indicating that severe wear occurred. In striking contrast with this, for the 5 wt % FL-GNP composite such abrasion grooves are almost completely absent. The corresponding wear volumes have been measured and the wear rates calculated (Fig. 5a). We found that wear rate measured for the 5 wt %FL-GNP sample ($5.9 \cdot 10^{-7}$ $mm^3$/N.m) is about twenty times lower than the wear rate of the monolithic $Si_3N_4$ ($1.2 \cdot 10^{-5}$ $mm^3$/N.m). This is a spectacular improvement of the wear resistance in full agreement with the optical microscopy images of the wear tracks. For comparison, composites with 3 wt % of GNP display a 14% reduction of wear rate while for the composites with 3 wt% of FL-GNP a 41% reduction has been measured.

The outstanding tribological performance achieved for 5 wt %FL-GNP composites can only be fully exploited in technological applications, if this does not come at the expense of the other mechanical properties. In order to clarify this we have measured the hardness, the three-point bending strength and the fracture toughness of the prepared composites on at least three samples. The results are summarized in Fig. 5b. Monolithic $S_3N_4$ samples display a Vickers hardness of 16.3 +/- 0.4 GPa, their three-point bending strength was found to be 549+/-23 MPa, while the fracture toughness 5.0 +/- 0.25MPa m$^{1/2}$. As it can be seen for the 3 wt% GNP samples the hardness and bending strength has been slightly decreased while the fracture toughness increased by about 12% in agreement with previous findings[9,10]. For 3wt% FL-GNP addition the hardness and flexural strength of the composite has already preserved the values of the monolithic $S_3N_4$, while the toughness further increased. Finally, for the tribologically best performing 5 wt% FL-GNP sample all the investigated mechanical properties have also been improved. This is most likely due to the fact that a better exfoliation rate leads to a more homogeneous dispersion, as thick GNPs can be regarded as larger aggregates of single layer graphene sheets. Nevertheless, the improvement of the mechanical properties even in the best case is rather moderate of order of 10 - 20%. This might be due to the rather weak interaction between the FL-GNPs and the $S_3N_4$ ceramic matrix, which on one hand is less efficient for improving the mechanical properties, on the other hand it is beneficial for improving the tribological performance allowing the FL-GNPs embedded in the matrix to be pulled out and incorporated into the tribo-film, giving rise to the self-lubricating nature of the composite.

In order to directly confirm that at the origin of the outstanding tribological properties of 5 wt% FL-GNP composite is indeed the formation of a FL-GNP tribo-film we have performed confocal Raman measurements in the wear tracks. Such Raman mapping has already proven very useful for the characterization of GNP distribution of sample surfaces and wear tracks[11,23]. A characteristic Raman map is shown in Fig. 6a. The red color symbolizes the presence of G-peak, while the brightness of the color indicates the intensity of the G peak. As can be seen the whole map is red, which means that in the wear track FL-GNPs form a continuous film, thought of varying thickness. Such maps have been acquired on various positions with similar results, indicating that all the investigated wear track areas are covered by a continuous FL-GNP tribo-film. Typical Raman spectra form the bright (thicker) and dark (thinner) areas of the map are shown in Fig. 6b. It is worth noting that even though the intensity ratio of the G and 2D peaks varies with location, both spectra acquired in the bright and dark areas display 2D peaks, characteristic to few-layer graphene[17,24,25]. Another important information that can be obtained from these spectra is the relatively low intensity of the D peak, indicating that even after the wear process the FL-GNPs preserve their high structural quality.

## 4. Conclusions

In summary, we have shown that the addition of graphene nanoplatelets with thickness approaching the two-dimensional limit to $Si_3N_4$ ceramics, remarkably improves the tribological performance of such ceramics. As compared to monolithic $Si_3N_4$, only 5wt% of few layer graphene nanoplatelet addition can improve the wear resistance by more than 20 times, while reducing the friction coefficient by 50%. By confocal Raman spectroscopic mapping of the wear tracks we have shown that this outstanding improvement in the tribological properties can be attributed to the formation of a continuous protecting and lubricating tribo-film consisting of FL-GNPs of high structural quality. The hardness, fracture toughness and bending strength of such composites are also improved compared to monolithic $Si_3N_4$. There are key technological advantages in developing highly wear-resistant and low-friction ceramic composite materials. Components produced from such materials enable the substantial reduction of losses during operation, as well as a significant increase of their durability.

**Acknowledgements**

O.T. was supported by OTKA grant PD 121368. Funding from the National Research Development and Innovation Office M-ERANET project „Graphene-ceramic composites for tribological application in aqueous environments" is also acknowledged. L.T. acknowledges the ERC Starting Grant, the "Lendület" programme of the Hungarian Academy of Sciences, and OTKA grant K10875.

# Figures and Tables

| Material | App. Density (g/cm$^3$) | α-Si$_3$N$_4$ (%) | β-Si$_3$N$_4$ (%) | ZrO$_2$ (%) | D Si$_3$N$_4$ (nm) |
|---|---|---|---|---|---|
| Si$_3$N$_4$ | 3.32 | 38,6 | 58,7 | 2,4 | 274 |
| Si$_3$N$_4$/3%GNP | 3.28 | 40,1 | 54,3 | 2,1 | 263 |
| Si$_3$N$_4$/3%FL-GNP | 3.3 | 60 | 33,2 | 1.1 | 240 |
| Si$_3$N$_4$/5%FL-GNP | 3.13 | 53.2 | 36,7 | 2.1 | 243 |

**Table 1.** The apparent density, phase composition, and average Si$_3$N$_4$ grain size of the composites.

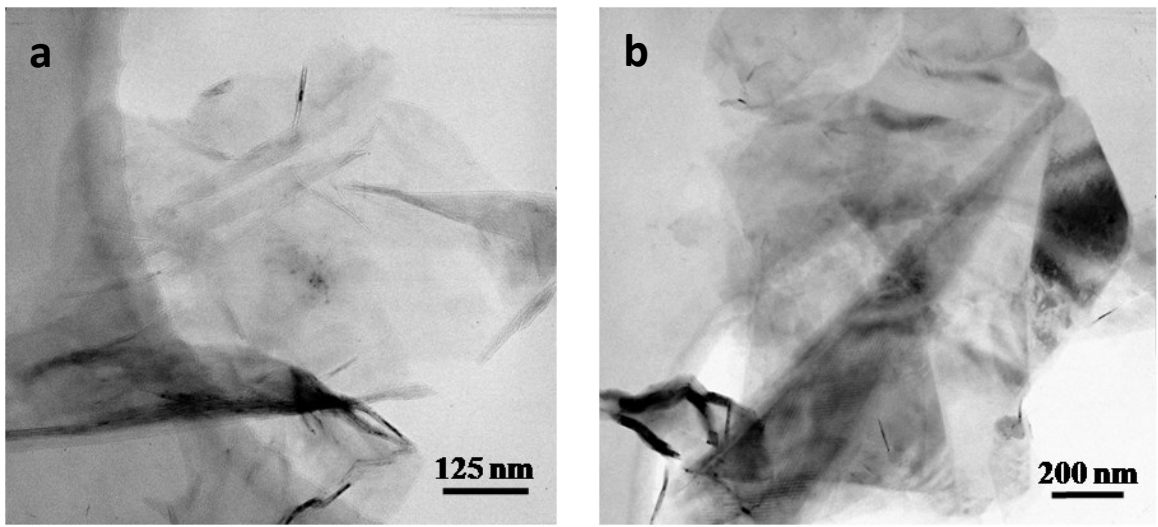

**Fig.1.** Transmission electron microscopy images of graphene nanoplatelets exfoliated by mechanical milling. Melamine addition to the milling process results in substantially thinner few-layer GNPs *(a)* as compared to those obtained by conventional exfoliation *(b)*.

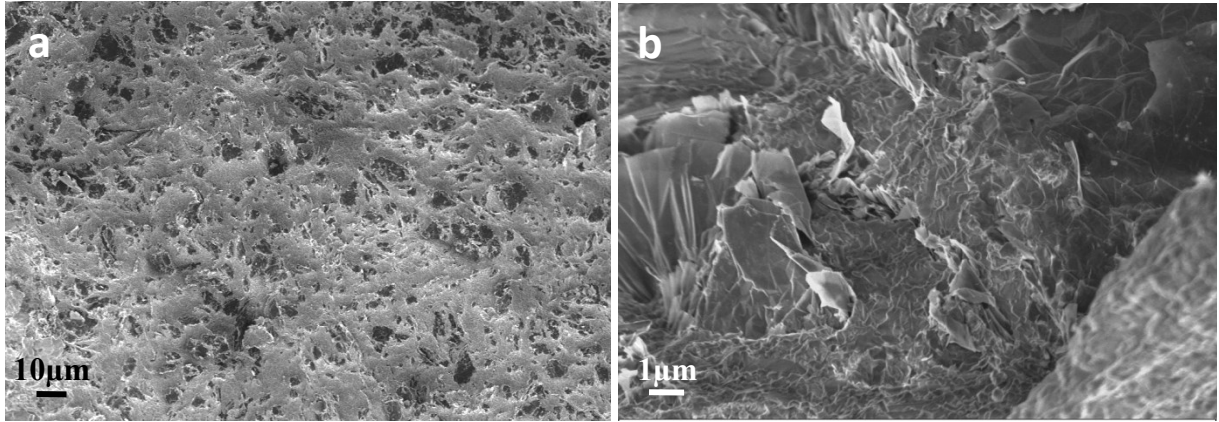

**Fig.2.** Scanning electron microscopy images of fracture surfaces of $Si_3N_4$ composites with 5wt% of FL-GNP addition. *a)* Lower magnification images reveal a relatively high surface density of micron-scale dark areas that can be identified in higher resolution SEM images *(b)* as few-layer graphene nanoplatelets oriented paralel to the surface.

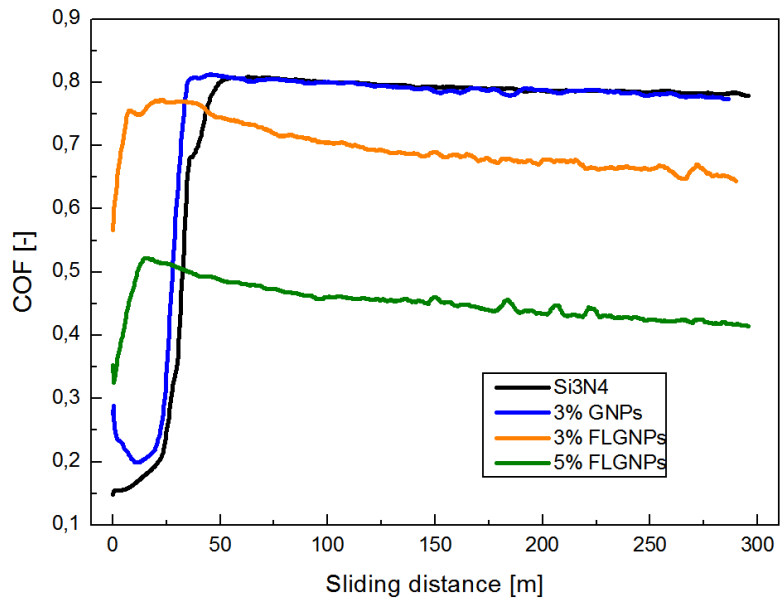

**Fig. 3.** Friction coefficient of various composites measured under dry sliding conditions with $Si_3N_4$ ball, at 5N loading, and 10 cm/s sliding speed. Composites with 5wt% of FL-GNP addition display a highly reduced friction coefficient.

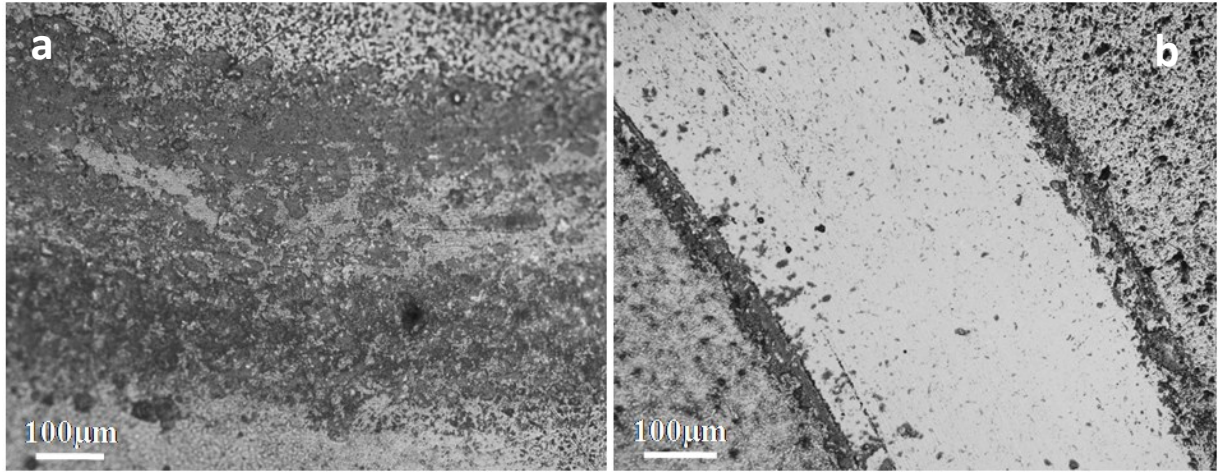

**Fig. 4.** Optical microscopy images of wear tracks for *a)* monolithic $Si_3N_4$ and *b)* $Si_3N_4$/5wt% FL-GNP composite. While abrasion groves are present in high density in monolithic samples, they are almost completely absent for the composites with 5 wt% of FL-GNP addition

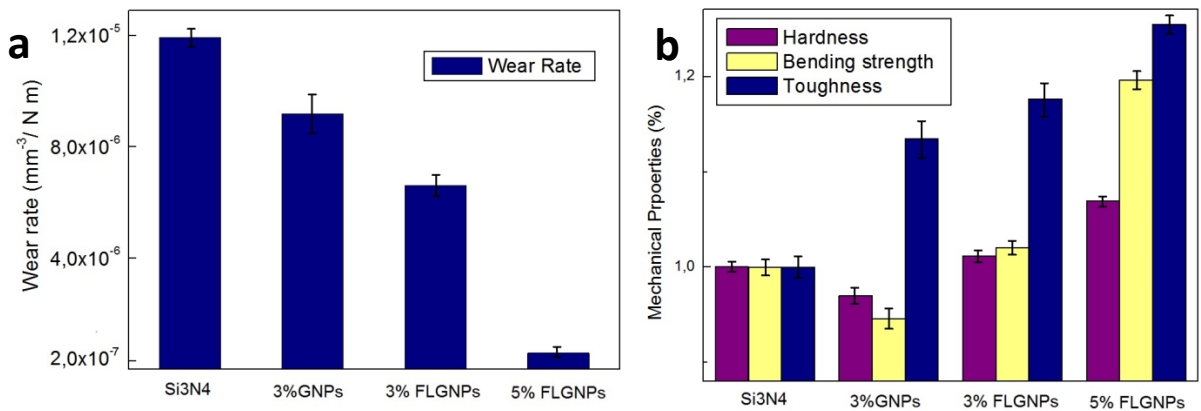

**Fig. 5.** a) Wear rate of various $Si_3N_4$ composites displaying a striking improvement of the wear resistance for composites with 5 wt% of FL-GNP addition. b) The Vickers hardness, three-point bending strength, and fracture toughness of $Si_3N_4$ composites normalized to the corresponding values of the monolithic $Si_3N_4$ sample to allow an easy comparison.

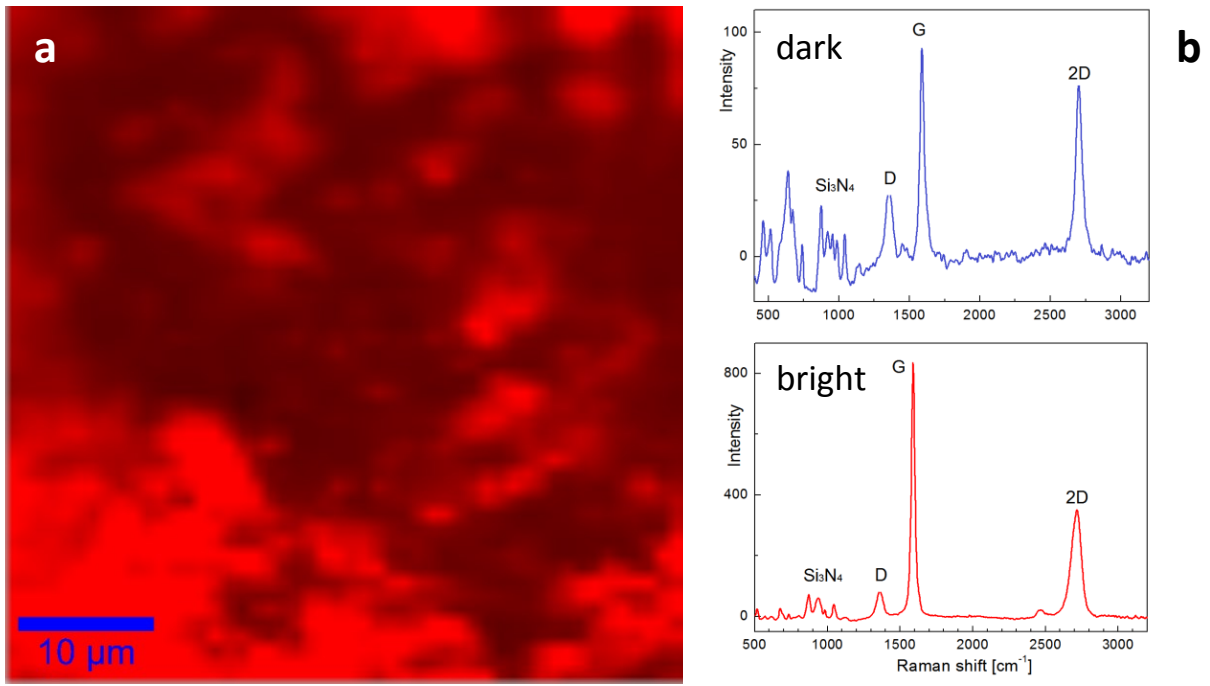

**Fig. 6.** a) False colored confocal Raman spectroscopy map of the G line of FL-GNPs recorded inside the wear track of a $Si_3N_4$ / 5 wt% FL-GNP sample. The red color indicates a high G line intensity. b) Characteristic Raman spectra inside the wear track, displaying the main lines of both the FL-GNPs and the $Si_3N_4$ matrix. A low D/G peak ratio indicates the highly preserved structural quality of FL-GNPs even after the wear process.

# References

[1] L.S. Walker, V.R. Marotto, M.A. Rafiee, N. Koratkar, E.L. Corral, Toughening in graphene ceramic composites, ACS Nano 5 (2011) 182–90.

[2] W. Bollmann, J. Spreadborough, Action of graphite as a lubricant, Nature 186 (1960) 29–30.

[3] H. J. Kim, S.-M. Lee, Y.-S. Oh, Y.-H. Yang, Y. S. Lim, D. H. Yoon, C. Lee, J.-Y. Kim, R. S. Ruoff, Unoxidized graphene/alumina nanocomposite: fracture- and wear-resistance effects of graphene on alumina matrix, Scientific Reports 4 (2014) 5176.

[4] H. Porwal, P. Tatarko, R. Saggar, S. Grasso, M. K. Mani, I. Dlouhý, J. Dusza, M. J. Reece, Tribological properties of silica–graphene nano-platelet composites, Ceramics international 40 (2014) 12067–12074.

[5] J. Llorente, B. Román-Manso, P. Miranzo, M. Belmonte, Tribological performance under dry sliding conditions of graphene/silicon carbide composites, Journal of the European Ceramic Society 36 (2016) 429–435.

[6] P. Hvizdos, J. Dusza, Cs. Balázsi, Tribological properties of $Si_3N_4$-graphene nanocomposites, J. Eur. Ceram. Soc. 33 (2013) 2359-2364.

[7] H. Li, Y. Xie, K. Li, L. Huang, S. Huang, B. Zhao, X. Zheng, Microstructure and wear behavior of graphene nanosheets-reinforced zirconia coating, Ceramics International 40 (2014) 12821–12829.

[8] C. Ramirez, F.M. Figueiredo, P. Miranzo, P. Poza, M.I. Osendi, Graphene nanoplatelet/silicon nitride composites with high electrical conductivity, Carbon 50 (2012) 3607–15.

[9] L. Kvetková, A. Duszová, P. Hvizdoš, J. Dusza, P. Kun, Cs. Balázsi, Fracture toughness and toughening mechanisms in graphene platelet reinforced Si3N4 composites, Scripta Materialia 66 (2012) 793–796.

[10] C. Ramirez, P. Miranzo, M. Belmonte, M. I. Osendi, P. Poza, S. M. Vega-Diaz, et al., Extraordinary toughening enhancement and flexural strength in Si3N4 composites using graphene sheets, 34 (2014) 161-169.

[11] M. Belmonte, C. Ramirez, J. Gonzalez-Julian, J. Schneider, P. Miranzo, M. I. Osendi, The beneficial effect of graphene nanofillers on the tribological performance of ceramics, Carbon 61 (2013) 431-435.

[12] P. Rutkowski, L. Stobierski, D. Zientara, L. Jaworska, P. Klimczyk, M. Urbanik, The influence of the graphene additive on mechanical properties and wear of hot-pressed Si3N4 matrix composites, Journal of the European Ceramic Society 35 (2015) 87–94.

[13] M. Maros, A.K. Németh, Z.Károly, E.Bódis, Z.Maros, O.Tapasztó, et al, Tribological characterisation of silicon nitride/multilayer graphene nanocomposites produced by HIP and SPS technology, Tribology International 93 (2016) 269-281.

[14] B. Lee, D. Lee, J. H. Lee, H. J. Ryu, S. H. Hong, Enhancement of toughness and wear resistance in boron nitride nanoplatelet (BNNP) reinforced $Si_3N_4$ nanocomposites, Scientific Reports 6 (2016) 27609.

[15] C. Lee, Q. Li, W. Kalb, X.Z. Liu, H. Berger, R.W. Carpick et al., Frictional characteristics of atomically thin sheets, Science 328 (2010) 76–80.

[16] Y. Hernandez V. Nicolosi, M. Lotya, F.M. Blighe, Z. Sun, S. De et al, High-yield production of graphene by liquid-phase exfoliation of graphite, Nature 3 (2008) 563.

[17] V. León, M. Quintana, M. A. Herrero, J. L. G. Fierro, A. de la Hoz, M. Prato, E. Vázquez, Few-layer graphenes from ball-milling of graphite with melamine, Chem. Commun. 47 (2011) 10936-10938.

[18] V. León, A. M. Rodriguez, P. Prieto, M. Prato, E. Vázquez, Exfoliation of graphite with triazine derivatives under ball-milling conditions: Preparation of few-layer graphene via selective noncovalent interactions, ACS Nano, 8 (2014) 563–571.

[19] P. Kun, F. Wéber, Cs. Balázsi, Preparation and examination of multilayer graphene nanosheets by exfoliation of graphite in high efficient attritor mill, Central European Journal of Chemistry 9 (2011) 47–51.

[20] G.R. Anstis, P. Chantikul, B.R. Lawn, D.B. Marshall, A critical evaluation or indentation techniques for measuring fracture toughness: I, direct crack measurements, J. Am. Ceram. Soc., 64 (1981) 533–538.

[21] P. Kapil, R. Rupesh, H. Cheng, Polymeric organo–magnesium complex for room temperature hydrogen physisorption. RSC Adv. 5 (2015) 10886-10891

[22] O. Tapasztó, L. Tapasztó, H. Lemmel, V.Puchy, J. Dusza, Cs. Balázsi, et al, High orientation degree of graphene nanoplatelets in silicon nitride composites prepared by spark plasma sintering, Ceramics International 42 (2016) 1002.

[23] C. Ramirez, M.I. Osendi, Characterization of graphene nanoplatelets-Si3N4 composites by Raman spectroscopy, Journal of the European Ceramic Society, 33 (2013) 471-477.

[24] A.C. Ferrari, J.C. Meyer, V. Scardaci, C. Casiraghi, M. Lazzeri, F. Mauri, et al. Raman spectrum of graphene and graphene layers. Physical Review Letters, 97 (2006) 187401

[25] A. Gupta, G. Chen, P. Joshi, S. Tadigadapa, P.C., Raman Scattering from High-Frequency Phonons in Supported *n*-Graphene Layer Films. Nano Letters, 6 (2006) 2667-2673